\documentclass[12pt,a4paper,twocolumn]{article}
\usepackage[T1]{fontenc}
\usepackage{lmodern}
\usepackage[margin=17mm,top=18mm,bottom=19mm,columnsep=7mm]{geometry}
\usepackage{amsmath,amssymb,graphicx,booktabs,array}
\usepackage[font=small,labelfont=bf,labelsep=period]{caption}
\usepackage{microtype}
\usepackage{titlesec}
\usepackage{enumitem}
\usepackage{stfloats}
\usepackage{flushend}
\usepackage{url}
\usepackage{xcolor}
\usepackage[hidelinks,breaklinks=true]{hyperref}

\hypersetup{
  pdftitle={Streaming instabilities and electron transport in electron transpiration cooling},
  pdfauthor={Wushun Zhang, Yinjian Zhao, Weixing Zhou}
}

\graphicspath{{figures/}}
\titleformat{\section}{\large\bfseries}{\thesection.}{0.5em}{}
\titleformat{\subsection}{\normalsize\bfseries}{\thesubsection.}{0.5em}{}
\titlespacing*{\section}{0pt}{12pt plus 3pt minus 2pt}{6pt}
\titlespacing*{\subsection}{0pt}{9pt plus 2pt minus 1pt}{4pt}

\newcommand{\NHe}{N_{\mathrm{He}}}
\newcommand{\units}[1]{\,\mathrm{#1}}
\newcommand{\figref}[1]{figure~\ref{#1}}
\newcommand{\tabref}[1]{table~\ref{#1}}
\newcommand{\secref}[1]{section~\ref{#1}}
\newcommand{\eqrefn}[1]{equation~\eqref{#1}}
\begin{document}

\twocolumn[
\begin{@twocolumnfalse}
\begin{center}
{\LARGE\bfseries
Streaming instabilities and electron transport in electron transpiration cooling
\par}

\vspace{6pt}
{\large
Wushun Zhang\textsuperscript{1,2},
Yinjian Zhao\textsuperscript{1,*}, and
Weixing Zhou\textsuperscript{1,2,*}\par}

\vspace{4pt}
{\small
\textsuperscript{1}School of Energy Science and Engineering,
Harbin Institute of Technology, Harbin 150001,
People's Republic of China\par
\textsuperscript{2}Zhengzhou Advanced Research Institute of
Harbin Institute of Technology, Zhengzhou 450000,
People's Republic of China\par}

\vspace{3pt}
{\small
\textsuperscript{*}Corresponding authors:\par
Yinjian Zhao (\href{mailto:zhaoyinjian@hit.edu.cn}{zhaoyinjian@hit.edu.cn})\par
Weixing Zhou (\href{mailto:zhouweixing@hit.edu.cn}{zhouweixing@hit.edu.cn})\par}

\end{center}

\vspace{3pt}
\noindent\textbf{Abstract.}
Electron transpiration cooling requires emitted electrons to travel through a surrounding plasma, where their relative motion with background electrons can drive streaming instabilities. We investigate these instabilities using particle-in-cell simulations with Monte Carlo collisions in a finite helium plasma. Lower neutral densities show pronounced wave activity, while higher densities give smoother electron distributions. A multifluid model constructed from simulated velocity distributions captures the frequency scale and decay direction in selected cases, but underestimates the separate driving and absorbing responses of electron populations. Distinguishing fast and slow emitted electrons and normalizing the field amplitude by the expanding two-peak length improves the low-wavenumber growth comparison, while discrepancies remain. Spatial growth estimates show that finite propagation distance also limits amplification. Strong late-time current fluctuations can coexist with nearly complete mean anode collection. These findings connect streaming instability to electron transport in the cooling model and show both the usefulness and limitations of a fluid description.

\vspace{5pt}

\vspace{12pt}
\end{@twocolumnfalse}
]

\section{Introduction}

Electron transpiration cooling (ETC) removes energy from a hot surface through thermionic electron emission and is being explored for hypersonic thermal protection~\cite{r3}. Laboratory studies have demonstrated cooling associated with electron emission in plasma environments~\cite{r6,r7}. The surrounding plasma influences how many emitted electrons escape the surface, return to it or reach a collector. Recent work on ETC circuits treats the emitting surface, plasma, collecting surface and internal electrical path as parts of one system~\cite{r5}. Understanding electron transport through the plasma is consequently part of assessing ETC operation.

Space charge near the emitting surface can restrict electron escape, while ions can modify the potential barrier and transmitted current~\cite{r10,r13}. Backflow provides a further constraint that depends on processes in the plasma interior and at the anode~\cite{r35}. Our previous full-gap kinetic study examined this broader transport problem for ETC, including emitted-electron backflow, potential redistribution and downstream collection~\cite{r14}. The present simulations also show electric-field fluctuations and evolving branches in electron phase space. These observations motivate examining the plasma dynamics that accompany electron transport across the gap.

Relative motion between emitted and background electrons provides a possible source of free energy for streaming instabilities. This mechanism has a direct precedent in thermionic plasmas: Levko and Krasik found electron two-stream instability and wave-induced electron heating in thermionic discharges, with the instability absent at higher gas pressure in their comparison~\cite{r36}. Kinetic studies of emissive sheaths also distinguish electron two-stream activity from oscillations involving trapped charge-exchange ions, and show that the potential fluctuations and velocity distributions depend on the domain and collisional conditions~\cite{r11,r12}. These results motivate identifying the electron populations that supply or absorb wave energy in the ETC transport setting, and examining how their response changes with neutral density.

A local fluid dispersion relation 
provides a framework 
for interpreting these fluctuations.
Finite boundaries and neutral collisions can alter beam-driven wave growth~\cite{r21,r22}. In the present system, the velocity distributions and the region containing distinct electron streams also evolve. A rate inferred from a spatial field projection can therefore reflect changes in both field strength and spatial structure. Its net value can also conceal large, opposing contributions from different populations. The comparison must examine how the model represents those populations and their energy exchange. For ETC, a further question is whether strong temporal fluctuations necessarily accompany poor mean electron collection.

We investigate these questions using electrostatic particle-in-cell (PIC) simulations with Monte Carlo collisions in a finite helium plasma. A neutral-density scan establishes how electron phase space and field fluctuations change under different collisional conditions. We then compare simulated wave evolution and population currents with a multifluid approximation constructed from the evolving velocity distributions, and examine late-time anode collection. Two distinctions emerge: agreement in a net wave rate need not imply agreement in the driving and absorbing responses, and strong current fluctuations need not imply poor mean collection. These results connect streaming dynamics to electron transport relevant to ETC while identifying the limits of the fluid description.

\section{Physical model and PIC simulations}
\label{sec:model}

\subsection{Plasma gap, emission and collisions}

\begin{figure*}[!tb]
\centering
\includegraphics[width=\textwidth]{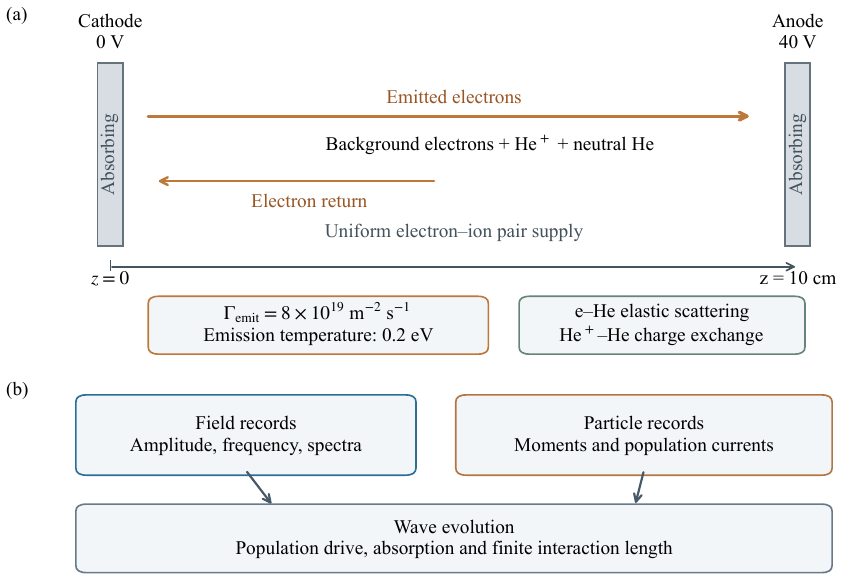}
\caption{Simulation model. (a) Helium plasma gap with absorbing electrodes, thermionic electron injection, uniform electron--ion pair replenishment and neutral collisions. (b) Field and particle records used to study wave evolution and population response.}
\label{fig:model}
\end{figure*}

The model is a finite helium plasma between an emitting cathode and an absorbing anode (\figref{fig:model}(a)). Electrostatic PIC simulations with Monte Carlo collisions (MCC) are implemented in WarpX with one spatial and three velocity dimensions (1D3V)~\cite{r24,r25}. The gap, bias, initial plasma state and emission temperature follow the full-diode benchmark used in our previous study~\cite{r14}. The axial coordinate $z$ runs from the cathode at 0 V to the anode at 40 V, with a separation $L_{\mathrm{gap}}=0.10\units{m}$. Both electrodes absorb incident particles. Emitted electrons, background electrons and He$^+$ ions are tracked separately, and their total charge determines the potential and axial field $E_z$ through Poisson's equation.

The cathode injects a prescribed electron number flux
$\Gamma_{\mathrm{emit}}=8\times10^{19}\units{m^{-2}\,s^{-1}}$
at $T_{\mathrm{emit}}=0.2\units{eV}$.
Emitted-electron velocities are sampled from a flux-weighted
half-Maxwellian distribution~\cite{r26}.
At each time step, 
ion losses to both electrodes 
are estimated
from the current particle positions and velocities.
An equal number 
of electron--ion pairs 
is then added 
at uniformly sampled positions 
within the gap.
The electron and ion velocities 
in each pair are drawn from
zero-drift Maxwellians 
at 3.0 and 0.15 eV, 
respectively,
using the initial background macroparticle weight.

Electron--He collisions use the Biagi isotropic elastic-scattering cross section~\cite{r28}. The He$^+$--He charge-exchange table is constructed as $\sigma_{\mathrm{cx}}=\sigma_{\mathrm{bs}}+\tfrac12\sigma_{\mathrm{iso}}$, where the backscattering and isotropic ion-scattering cross sections are taken from the Phelps data in LXCat~\cite{r29}. Collision rates depend on energy and relative velocity~\cite{r30}; the neutral bath is at 300 K. These prescribed emission, replenishment and collision processes define the thermionic transport model considered here.

\subsection{Numerical settings and density scan}

The initial plasma density is $n_0=5\times10^{14}\units{m^{-3}}$, with electron and ion temperatures of 3.0 and 0.15 eV. The baseline calculation uses 4096 cells and 1000 initial macroparticles per cell for each background species.
With $\Delta z=L_{\mathrm{gap}}/4096$,
the time step is
$\Delta t=0.2\Delta z/v_{\mathrm{te}}
=6.722\times10^{-12}\,\mathrm{s}$,
where $v_{\mathrm{te}}=\sqrt{eT_e/m_e}$ is the initial
electron thermal speed.
Here $e$ is the elementary charge and $m_e$ the electron mass.
The initial Debye length spans 23.6 cells.

The baseline neutral densities are $\NHe=\{0.05,0.1,0.5,1,1.5,2,2.9,5\}\times10^{21}\units{m^{-3}}$. Additional intermediate densities are used for the decay comparison below. Plasma and emission inputs remain fixed, so the scan changes the collisional conditions without changing the prescribed electron supply.

Grid, time-step, particle-number and random-seed checks, following the verification approach discussed in~\cite{r27}, cover 2048--8192 cells, time steps down to one quarter of baseline, and 500--2000 particles per cell. They assess the early-decay population balance discussed in \secref{sec:balance}. Figure~\ref{fig:model}(b) shows how the field records provide wave amplitudes and frequencies and how the particle records provide moments and population currents.

\section{Wave and electron evolution in the PIC simulations}
\label{sec:observations}

\begin{figure*}[!tb]
\centering
\includegraphics[width=\textwidth]{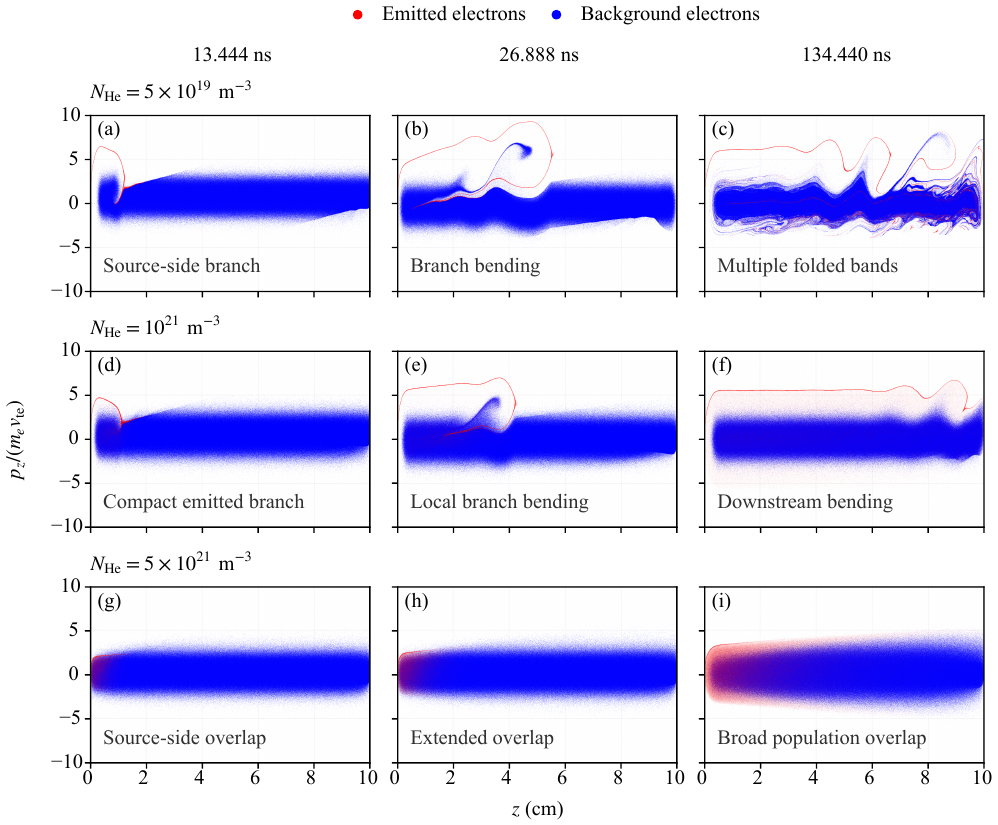}
\caption{Early electron phase-space evolution. Rows: $\NHe=5\times10^{19}$, $10^{21}$ and $5\times10^{21}\units{m^{-3}}$. Columns: 13.444, 26.888 and 134.440 ns. Red and blue denote emitted and background electrons, respectively. The normalized momentum scale is common to all panels.}
\label{fig:early}
\end{figure*}

\subsection{Electron phase space across the density scan}

\begin{figure*}[!tb]
\centering
\includegraphics[width=\textwidth]{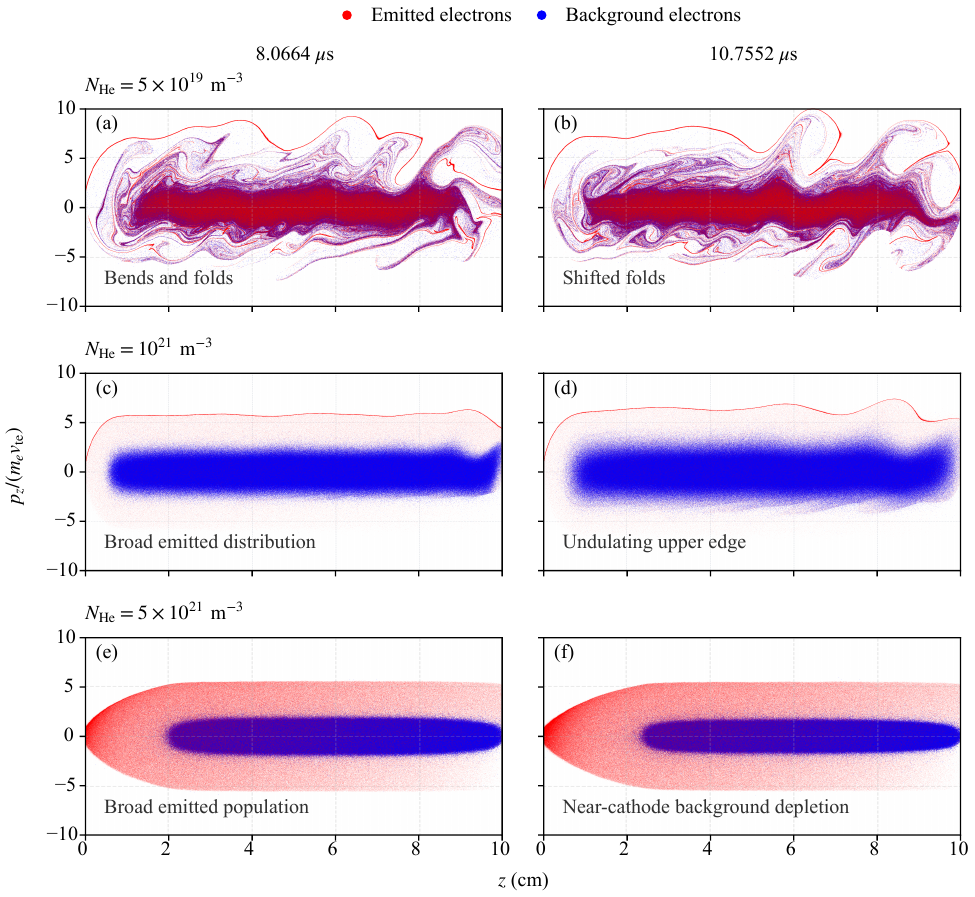}
\caption{Late electron phase space at 8.0664 and 10.7552 $\mu$s. Rows: $\NHe=5\times10^{19}$, $10^{21}$ and $5\times10^{21}\units{m^{-3}}$. Red denotes emitted electrons and blue background electrons. Axes and momentum normalization match \figref{fig:early}.}
\label{fig:late}
\end{figure*}

The broad picture is established by comparing the lowest, an intermediate and the highest neutral density. Figure~\ref{fig:early} shows electron phase space at 13.444, 26.888 and 134.440 ns. Red denotes emitted electrons and blue denotes background electrons. All panels use the same normalized axial momentum $p_z/(m_ev_{\mathrm{te}})$, so their velocity structures can be compared directly.

At $\NHe=5\times10^{19}\units{m^{-3}}$, emitted electrons form a source-side branch that subsequently bends and develops multiple folded structures (\figref{fig:early}(a)--(c)). Local bends also occur at $10^{21}\units{m^{-3}}$, although the later structure is smoother (\figref{fig:early}(d)--(f)). At $5\times10^{21}\units{m^{-3}}$, emitted and background distributions overlap more strongly (\figref{fig:early}(g)--(i)).
With the emission 
and initial plasma conditions 
held fixed, 
the electron phase-space structures 
vary markedly 
with neutral gas density.

Figure~\ref{fig:late} extends 
this overview to 8.0664 and
10.7552 $\mu$s 
for the same three cases.
At the lowest density, bent and folded branches persist while their positions change (\figref{fig:late}(a,b)).
At the intermediate density, the distributions become broader
and smoother, with spatial modulation still visible at the emitted-electron boundary (\figref{fig:late}(c,d)).
At the highest density, both snapshots show a broad emitted
distribution and depleted background electrons near the cathode (\figref{fig:late}(e,f)).

These snapshots establish the evolution that the following analysis seeks to explain. The relative electron motion supplies free energy for streaming activity, but phase-space appearance alone does not determine a mode's growth rate. We therefore turn first to the electric field, and then to a model built from the simulated velocity distributions. The late-time collection statistics are considered in \secref{sec:transport}.

\begin{figure*}[!tb]
\centering
\includegraphics[width=\textwidth]{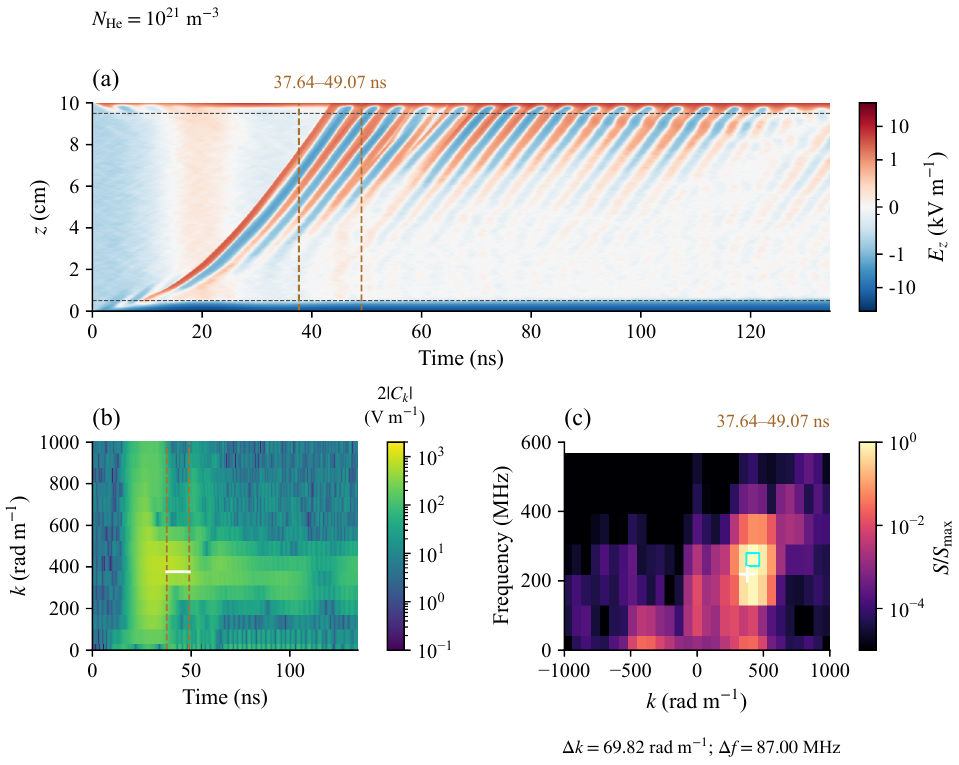}
\caption{Electric-field structure at $\NHe=10^{21}\units{m^{-3}}$.
(a) Axial electric field; dashed lines mark the 0.5--9.5 cm
observation region and 37.64--49.07 ns decay interval.
The colour scale is linear within $\pm0.5\units{kV\,m^{-1}}$
and logarithmic outside.
(b) Amplitude $2|C_k|$; the white line marks the selected
wavenumber $k=376.991\units{rad\,m^{-1}}$.
(c) Spatiotemporal spectrum over the same interval, including $k=0$.
The bin spacings are $\Delta k=69.82\units{rad\,m^{-1}}$
and $\Delta f=87.00\units{MHz}$.
The square marks the spectral peak ($\sim261$ MHz),
and the cross marks the PIC phase-fit frequency
(218.541 MHz) at the selected $k$.}
\label{fig:field}
\end{figure*}

\subsection{Electric-field structure
and wave diagnostics}
\label{sec:field}

At $\NHe=10^{21}\units{m^{-3}}$,
propagating field striations 
develop 
as emitted electrons 
spread downstream (\figref{fig:field}(a)). 
The applied bias accelerates 
emitted electrons 
relative to the background. 
The field record 
shows 
an evolving wave pattern, 
providing a more direct measure 
of 
the fluctuations 
than the phase-space snapshots alone.

To quantify the field, 
we use the common 
interior region $0.5\le z<9.5\units{cm}$.
Omitting 0.5 cm 
next to each electrode 
reduces the influence 
of strong near-electrode fields 
while 
retaining a 9 cm comparison region.

At each time $t$, we calculate the arithmetic mean of the axial
electric field over the selected region as
$B(t)=N^{-1}\sum_{j=1}^{N} E_z(z_j,t)$, where $N$ is the number
of spatial sampling points in that region. We subtract $B(t)$ from
each field value before applying the periodic Hann window~\cite{r31}.
For a given axial wavenumber $k$, 
the complex
electric-field coefficient $C_k(t)$ 
and the
corresponding field amplitude 
$A_k(t)$ are defined as
\begin{equation}
\label{eq:projection}
\begin{aligned}
C_k(t)&=\frac{\sum_j w_j[E_z(z_j,t)-B(t)]e^{-ik(z_j-z_0)}}{W},\\
W&=\sum_jw_j,\qquad A_k=2|C_k|.
\end{aligned}
\end{equation}
Here, 
$E_z(z_j,t)$ is 
the axial electric field
at the sampling position 
$z_j$,
$z_0=0.5\,\mathrm{cm}$ 
is the start of 
the observation region, 
and $w_j$ is 
the periodic Hann window weight
at $z_j$.
The sum runs over 
all sampling points 
within the region,
and $W$ is the sum 
of the weights.

Figure~\ref{fig:field}(b) shows how $A_k$ varies with time and wavenumber. The white line marks the wavenumber used for the decay fit.
The phase of $C_k$ is unwrapped 
to remove $2\pi$ jumps 
and then fitted linearly in time.
Dividing 
the fitted phase slope 
by $-2\pi$ gives
the signed frequency $f$.
The oscillation frequency 
is $|f|$.
A linear least-squares fit 
to $\ln A_k$ 
gives a net rate
$\gamma_{\mathrm{fit}}$: 
positive values indicate 
growth and 
negative values decay.
In both fits, 
the slope and intercept 
are determined 
from the data.
The coefficient 
of determination 
$R^2$ 
describes how closely 
the log amplitude 
follows 
a straight line, 
or equivalently 
how closely 
the amplitude 
follows a single exponential.
Each comparison 
below specifies the time interval 
used for fitting.

The spatiotemporal spectra 
also use 
a periodic Hann window
in time. 
The wavenumber 
and frequency bin spacings 
are
$\Delta k=2\pi/(N_z\Delta z)$ 
and
$\Delta f=1/(N_t\Delta t_f)$, 
where $N_z$ 
and $N_t$ 
are the
numbers 
of spatial and temporal samples, 
respectively, 
and
$\Delta t_f$ 
is the field-output interval.
For \figref{fig:field}(c), 
the frequency-bin spacing 
is
approximately 87 MHz. 
The phase fit 
over the same time interval
gives 
an oscillation frequency 
of 218.54 MHz, 
which is used
for the quantitative comparison 
with theory.

For the intermediate-density cases, we identify a strong field
component near the electron plasma frequency from the full-gap
(10 cm) spectrum. The wavenumber selected for each case is then
used to project the PIC field over the 0.5--9.5 cm interior region
and to solve the theoretical dispersion relation.
For the lowest- and highest-density cases, we examine how separating
the emitted electrons affects growth or decay at selected wavenumbers
on the interior-grid spacing, $k_m=m\Delta k$, where $m$ is an
integer and $\Delta k=69.821\units{rad\,m^{-1}}$.

\section{Interpreting the waves 
with a fluid approximation}
\label{sec:theory}

\begin{figure*}[tb]
\centering
\includegraphics[width=\textwidth]{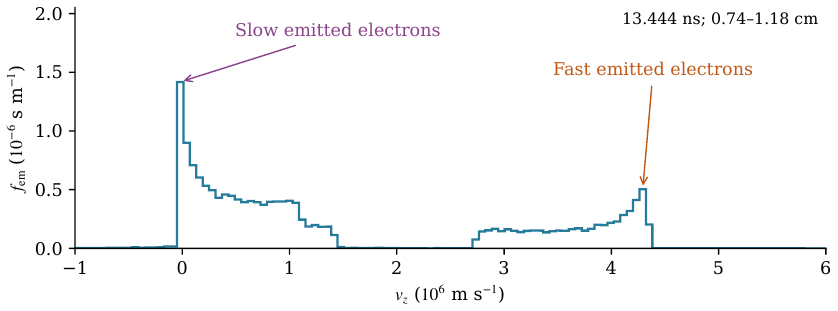}
\caption{Emitted-electron velocity distribution at $\NHe=5\times10^{19}\units{m^{-3}}$, 13.444 ns, over 0.74--1.18 cm. The particle-weighted histogram has unit integral. Its two peaks identify slow and fast emitted electrons; background electrons and ions provide the other populations in the fluid description.}
\label{fig:distribution}
\end{figure*}

\subsection{From velocity distributions 
to a dispersion relation}
\label{sec:dispersion}

The emitted population 
does not 
remain 
a single narrow stream. 
Fast and slow velocity 
peaks occur 
during the evolution 
of all neutral-density cases. 
Figure~\ref{fig:distribution} 
shows an example 
at the lowest density, 
where the two peaks 
are readily distinguished. 
This structure 
motivates 
two descriptions 
of the same particles. 
A three-population model 
merges 
all emitted electrons 
and treats 
background electrons 
and ions 
as the other two populations. 
A four-population model separates 
fast and slow emitted electrons,
retaining a distinct drift 
and pressure response for each.

When fast 
and slow emitted electrons 
are combined 
into one population,
the difference 
between their mean velocities 
contributes to 
the combined velocity spread.
Separating them 
at the midpoint 
between 
the two local velocity peaks
retains the mean velocity 
and velocity spread 
of each population.
For each case, 
both models 
use the same particle records 
and spatial bins.
Splitting 
the emitted electrons 
into fast 
and slow populations 
preserves 
their total number density 
and axial current density 
in each bin.
The three-population model 
is used 
to compare 
theoretical frequencies 
and decay rates 
with PIC results 
at intermediate neutral densities. 
At the lowest 
and highest densities, 
we compare 
the three- 
and four-population 
models 
to examine 
how separating 
fast 
and 
slow 
emitted electrons 
affects 
the predicted 
growth or decay rates.

For each population $s$, 
with charge $q_s$ 
and mass $m_s$,
we calculate the density $n_s$, 
axial drift velocity $u_s$,
and axial temperature
$T_s=m_s\langle(v_z-u_s)^2\rangle_s/e$
in each spatial bin, 
with $T_s$ expressed in eV.
Here, 
$\langle\cdot\rangle_s$ 
denotes an average 
over
population $s$, 
weighted 
by the 
macroparticle statistical weights.

For each population, 
the regional drift velocity 
and temperature 
are
averages 
of the corresponding bin values,
weighted 
by the total particle
weight 
in each bin. 
The regional mean number density 
is the total particle
weight 
in the region divided 
by its length.

Assuming stationary helium atoms, 
the fluid model 
uses a mean collision
frequency of
$\nu_s=\NHe\langle\sigma_s(\epsilon_{\mathrm{kin}})|\mathbf{v}|\rangle_s$
for population $s$. 
For each particle, 
the cross section 
at its kinetic
energy is obtained 
by linear interpolation 
of the tabulated data
specified in \secref{sec:model}.

At each sampled time, 
the particle moments 
define a locally uniform
background 
for the linear analysis. 
The particle supply rate 
is held
fixed 
as the density and velocity fluctuate.

We consider perturbations proportional 
to $\exp[i(kz-\omega t)]$,
where 
$\omega=\omega_r+i\gamma$,
$f=\omega_r/(2\pi)$ is 
the signed frequency,
and $\gamma$ is the growth rate. 
For population $s$, 
$\Omega_s=\omega-ku_s$ is 
the complex angular
frequency 
in the frame 
moving 
with its axial drift velocity $u_s$.
The first two equations 
express particle number 
conservation 
and axial
momentum balance 
for each population. 
The third is Poisson's equation,
which relates 
the axial electric field perturbation 
to the net charge
density perturbation 
from all populations~\cite{r17,r32}:
\begin{equation}
\label{eq:fluid}
\begin{aligned}
-i\Omega_s\delta n_s+ikn_s\delta u_s&=0,\\
(\nu_s-i\Omega_s)\delta u_s&=\frac{q_s}{m_s}\delta E-\frac{ik}{m_sn_s}\delta P_s,\\
ik\delta E&=\frac{1}{\epsilon_0}\sum_s q_s\delta n_s.
\end{aligned}
\end{equation}
Here 
$\delta n_s$, $\delta u_s$, 
and $\delta P_s$ 
are the perturbations
in number density, 
axial drift velocity, 
and axial pressure 
of population
$s$. 
The axial electric field 
perturbation is $\delta E$, 
and
$\epsilon_0$ 
is the vacuum permittivity. 
In the momentum equation, 
$\nu_s$ sets the rate 
at which collisions
damp the axial velocity perturbation~\cite{r22}.

We close the pressure response 
with $\delta P_s=m_sc_s^2\delta n_s$, where $c_s^2=\eta_s eT_s/m_s$. 
The isothermal model 
M1 uses $\eta_e=1$. 
Model M3 uses $\eta_e=3$, 
corresponding to 
one-dimensional adiabatic 
electron pressure 
with neglected 
heat flux~\cite{r32,r19}. 
For the electron-frequency 
branches considered here, 
the model retains
ion inertia 
and sets $\eta_i=0$, 
so that $\delta P_i=0$.

Eliminating the velocity perturbation 
gives
\begin{equation}
\label{eq:density}
\delta n_s=\frac{ikn_sq_s\delta E/m_s}{\Omega_s(\Omega_s+i\nu_s)-k^2c_s^2}.
\end{equation}
Substitution 
into Poisson's equation 
gives the dispersion relation
\begingroup
\small
\begin{equation}
\label{eq:dispersion}
\begin{aligned}
D(\omega,k)&=1-\sum_s\frac{\omega_{ps}^2}{\Omega_s(\Omega_s+i\nu_s)-k^2c_s^2}=0,\\
\omega_{ps}^2&=\frac{n_sq_s^2}{\epsilon_0m_s}.
\end{aligned}
\end{equation}
\endgroup
Each denominator 
combines drift, 
collisional relaxation 
and pressure response.
The charge-density perturbations 
of all populations 
determine
$\delta E$ 
through Poisson's equation.
This electric-field perturbation 
acts on every population.

For a given real wavenumber $k$, 
the three- 
and four-population models
yield six and eight candidate roots 
for $\omega$, 
respectively.
For each pressure closure 
(M1 or M3), 
the roots are computed separately
and checked 
by substitution 
into the original dispersion relation.
Each verified root 
represents 
a collective mode 
with angular frequency
$\omega_r=\operatorname{Re}\omega$ 
and temporal growth rate
$\gamma=\operatorname{Im}\omega$.

For the intermediate-density cases, 
we track a theoretical branch 
to compare
its decay rate 
with that measured 
in the PIC simulation. 
At the first sampling
time, 
we select a root 
whose frequency 
has the same sign 
and approximately
the same magnitude 
as the PIC frequency. 
At each subsequent step, 
we pair
the old and new roots 
to minimize the total distance 
between paired roots
in the complex frequency plane. 
The new root paired 
with the previously
selected root continues the branch. 
For the lowest- 
and highest-density
cases, 
we test 
whether 
each sampled background 
supports a growing mode 
at
the selected wavenumber $k$. 
We therefore 
select the root 
with the largest
temporal growth rate $\gamma=\operatorname{Im}\omega$. 
A positive maximum
gives the fastest growth rate, 
while a negative maximum 
means that all modes
decay.

\subsection{A first test: continuous wave decay}
\label{sec:decay}

\begin{figure*}[!tb]
\centering
\includegraphics[width=\textwidth]{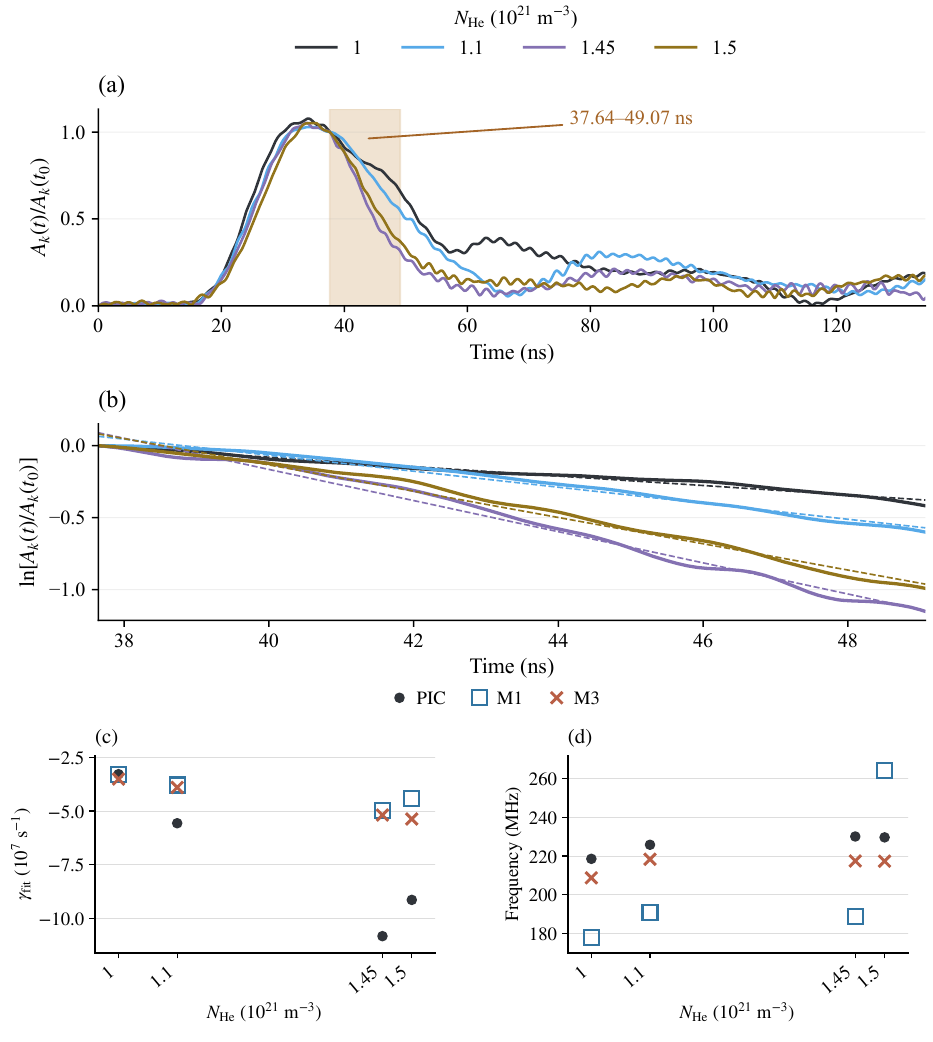}
\caption{Wave decay at four intermediate neutral
densities. (a) PIC field amplitudes in the 0.5--9.5~cm region.
Shading marks the common decay interval (37.643--49.071~ns).
(b) Log amplitudes $\ln[A_k(t)/A_k(t_0)]$ and their linear fits
(dashed), with $t_0=37.643$~ns.
(c) Fitted temporal growth rates and (d) frequencies over the same
interval for PIC and the three-population model with the M1 and M3
pressure closures. Negative growth rates indicate decay.}
\label{fig:decay_scan}
\end{figure*}

Figure~\ref{fig:decay_scan}(a) shows the field amplitudes
for the four intermediate neutral densities,
$\NHe=\{1,1.1,1.45,1.5\}\times10^{21}\units{m^{-3}}$.
In all four cases, the amplitude peaks and then decays continuously
from 37.643 to 49.071~ns. We use this common decay interval to compare
the frequencies and decay rates measured in the PIC simulations with
those predicted by the three-population model.
The PIC log amplitudes vary nearly linearly with time
over this interval, with $R^2=0.982$--$0.986$
(\figref{fig:decay_scan}(b) and \tabref{tab:fit}).

\begin{table}[!tb]
\caption{Fixed wavenumbers and PIC log-amplitude fit quality
over the common decay interval.}
\label{tab:fit}
\centering
\small
\begin{tabular}{ccc}
\toprule
$\NHe\ (10^{21}\units{m^{-3}})$ &
$k\ (\mathrm{rad\,m^{-1}})$ &
$R^2$\\
\midrule
1    & 376.991 & 0.983\\
1.1  & 439.823 & 0.982\\
1.45 & 439.823 & 0.986\\
1.5  & 439.823 & 0.985\\
\bottomrule
\end{tabular}
\end{table}

At the measured PIC frequencies of 218.5--230.1~MHz, the 11.428~ns
decay interval spans about 2.5--2.6 cycles. For each pressure closure,
the theory gives an instantaneous growth rate and angular frequency
at each sampling time. We integrate these values along the tracked
branch to obtain the changes in log amplitude and phase. We then apply
the same linear fits used for the PIC data at the same sampling times.
The log-amplitude slope gives the fitted theoretical growth rate for
the full interval. The fitted phase slope equals $-2\pi f$, where $f$
is the signed theoretical frequency.

The PIC amplitude decays more rapidly as $\NHe$ increases
from $10^{21}$ to $1.45\times10^{21}\units{m^{-3}}$, then more slowly
at $1.5\times10^{21}\units{m^{-3}}$ (\figref{fig:decay_scan}(c)).
Both M1 and M3 predict decay in all four cases.
At $10^{21}\units{m^{-3}}$, M1 gives the closer decay rate, while M3
overestimates the magnitude of the PIC decay rate by 7.0\%.
At the other three densities, M3 gives the closer decay rate, although
it underestimates the decay magnitude by 29.9--52.1\%.
M3 gives the closer frequency at all four densities
(\figref{fig:decay_scan}(d)). Relative to the PIC frequencies, the
differences are 3.3--5.5\% for M3 and 15.1--18.5\% for M1.

\begin{figure*}[!tb]
\centering
\includegraphics[width=0.96\textwidth]{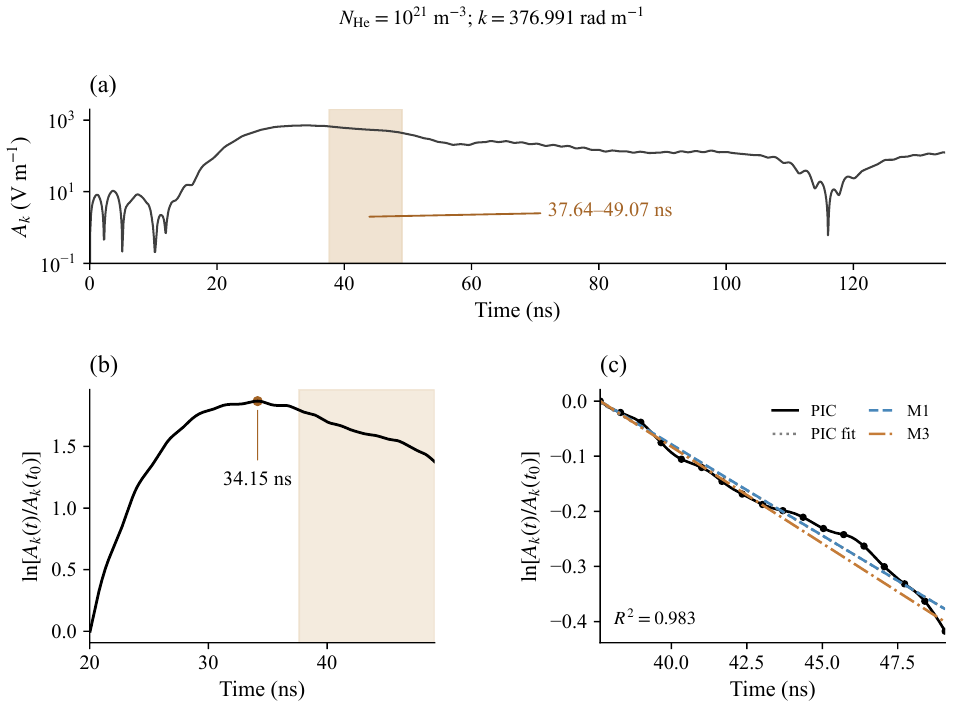}
\caption{Field amplitude at $\NHe=10^{21}\units{m^{-3}}$
and $k=376.991\units{rad\,m^{-1}}$.
(a) Full history, with 37.64--49.07~ns shaded.
(b) Growth and turnover at 34.15~ns.
(c) Comparison of PIC decay with the three-population
M1 and M3 models at the 18 particle-output times.
Panels (b) and (c) show $\ln[A_k(t)/A_k(t_0)]$,
where $t_0$ is the start of the respective time interval.
The reported $R^2$ is the coefficient of determination
for the linear fit to the PIC log amplitude.}
\label{fig:decay_detail}
\end{figure*}

The $10^{21}\units{m^{-3}}$ case has the smallest M3 decay-rate
discrepancy and is examined in detail in \figref{fig:decay_detail}
and \tabref{tab:rates}. Figure~\ref{fig:decay_detail}(a) shows the
full amplitude history, with the fitted decay interval highlighted.
The amplitude peaks at 34.15~ns
(\figref{fig:decay_detail}(b)), before the common decay interval begins.
At the selected wavenumber, the PIC phase fit gives 218.541~MHz,
marked by the cross in \figref{fig:field}(c).

\begin{table}[!tb]
\caption{Frequency and net rate over the same decay window
for the detailed case in \figref{fig:decay_detail}(c).}
\label{tab:rates}
\centering
\small
\setlength{\tabcolsep}{3pt}
\begin{tabular}{lrr}
\toprule
Method & $f$ (MHz) & $\gamma_{\mathrm{fit}}\ (10^7\units{s^{-1}})$\\
\midrule
PIC                   & 218.541 & $-3.279$\\
Three populations, M1 & 178.037 & $-3.302$\\
Three populations, M3 & 208.742 & $-3.508$\\
\bottomrule
\end{tabular}
\end{table}

Figure~\ref{fig:decay_detail}(c) compares the M1 and M3
amplitude histories with the fitted PIC decay. In this case, M1 gives
the closer decay rate, whereas M3 gives the closer frequency
(4.5\% below the PIC value, \tabref{tab:rates}).
Because M3 gives the closer frequency at all four densities, we use
it as the reference closure for the population-response comparison
that follows.

\begin{table*}[!tb]
\caption{Population contributions over 37.64--49.07~ns
in the detailed decay case.
Positive and negative values indicate driving and absorption,
respectively. All entries are in $10^8\units{s^{-1}}$.}
\label{tab:balance}
\centering
\small
\begin{tabular}{lrr}
\toprule
Contribution & PIC & M3\\
\midrule
Emitted electrons & $+2.0057$ & $+0.0470$\\
Background electrons & $-2.3312$ & $-0.3978$\\
Ions & $+1.94\times10^{-5}$ & $+4.33\times10^{-5}$\\
\midrule
Net rate from population currents & $-0.3255$ & $-0.3508$\\
Net rate from amplitude fit & $-0.3279$ & $-0.3508$\\
\bottomrule
\end{tabular}
\par\vspace{4pt}
\begin{minipage}{0.94\textwidth}
\footnotesize
In the PIC column, the emitted-electron row combines the contributions
from fast ($+2.1675$) and slow ($-0.1617$) emitted electrons.
The penultimate row gives the sum of $g_s$ for emitted electrons,
background electrons, and ions. The final row gives the rate fitted
to the field-amplitude history. For PIC, we use the measured field
amplitude; 
for M3, we integrate the instantaneous model growth rate over time
to obtain $\ln[A(t)/A(t_0)]$ and report the slope of its linear fit
over the decay interval.
\end{minipage}
\end{table*}

\subsection{Population drive, absorption and pressure response}
\label{sec:balance}

A net decay rate does not reveal the separate driving and
absorbing responses of individual populations.
We therefore examine population currents~\cite{r20,r33}.

In PIC, 
the axial current density 
of population $s$ 
is $J_s=q_s n_s u_s$.
At each time, 
we evaluate 
the axial electric field 
$E_z$ at each current bin centre 
by linear interpolation 
between the field values 
at the two grid points 
that bracket the centre. 
The current density 
and electric field 
are thus sampled 
at the same positions. 
Over $0.5$--$9.5$~cm,
we subtract the arithmetic mean 
of $J_s$ 
from the current density 
and
the arithmetic mean $B(t)$ 
of the interpolated field 
from the electric field. 
We then 
apply the Hann projection 
in equation~\eqref{eq:projection}
to both mean-subtracted quantities 
at the same wavenumber $k$. 
This gives 
the complex current coefficient 
$\mathcal{L}_k[J_s]$ 
and the
complex electric-field coefficient 
$C_k$.

The complex current response 
of population $s$ 
at wavenumber $k$ is
$H_s^{\mathrm{PIC}}=\mathcal{L}_k[J_s]/C_k$.
Its magnitude, 
$\lvert H_s^{\mathrm{PIC}}\rvert
=\lvert\mathcal{L}_k[J_s]\rvert/\lvert C_k\rvert$, 
is the ratio of
the current-density amplitude 
to the electric-field amplitude.
Its argument, 
$\phi_s=\arg(H_s^{\mathrm{PIC}})$, 
is the phase 
of the
current relative to the field, 
given by
$\phi_s=\arg(\mathcal{L}_k[J_s])-\arg(C_k)$.

The total axial current density is the sum $\sum_s J_s$ over all
particle populations.
Where this current density increases along $z$, more
charge flows out than flows in, so the local charge density decreases.
For the total charge density $\rho=\sum_s q_s n_s$, this balance is
expressed by the continuity equation
$\partial_t\rho=-\partial_z(\sum_s J_s)$.

Gauss's law,
$\epsilon_0\,\partial_z E_z=\rho$, relates the charge
density to the axial electric field. 
Differentiating Gauss's law 
with respect to 
time and 
substituting for 
$\partial_t\rho$ 
gives
$\partial_z(\epsilon_0\,\partial_t E_z+\sum_s J_s)=0$.
The quantity in parentheses 
is therefore spatially uniform 
at each time. 
Subtracting its spatial mean 
removes this uniform term. 
Applying the same Hann projection 
then gives
$\epsilon_0\,dC_k/dt+\sum_s\mathcal{L}_k[J_s]=0$.
Moving the current terms 
to the right-hand side 
and dividing by
$\epsilon_0 C_k$ yields
$(1/C_k)\,dC_k/dt=-(1/\epsilon_0)\sum_s H_s^{\mathrm{PIC}}$.

The real part of $(1/C_k)\,dC_k/dt$ 
gives the logarithmic amplitude
growth rate $d\ln\lvert C_k\rvert/dt$, while its imaginary part gives
the phase rate $d\arg(C_k)/dt$. One cycle corresponds to $2\pi$ radians.
Under the convention $e^{i(kz-\omega t)}$, a positive frequency gives
a decreasing phase, so the signed frequency is
$f=-(1/2\pi)\,d\arg(C_k)/dt$.

For the PIC frequency 
in Table~\ref{tab:rates}, 
we use $C_k$ from 
the original field grid. 
We fit a straight line 
to its unwrapped phase 
at the particle-output times 
within the decay interval.
The oscillation frequency 
$\lvert f\rvert$ 
is the absolute value
of the fitted slope divided 
by $2\pi$.

Taking the real part 
therefore gives the contribution $g_s$ of each
population to the field-amplitude growth rate. 
Summing these
contributions gives the net rate,
\begin{equation}
\label{eq:current}
\begin{aligned}
g_s
  &= -\frac{1}{\epsilon_0}
     \operatorname{Re}\!\left(\frac{\mathcal{L}_k[J_s]}{C_k}\right)
   = -\frac{\lvert H_s^{\mathrm{PIC}}\rvert}{\epsilon_0}\cos\phi_s, \\
\frac{d\ln\lvert C_k\rvert}{dt}
  &= \sum_s g_s .
\end{aligned}
\end{equation}
Here, 
$g_s$ has units of $\mathrm{s}^{-1}$. 

A positive $g_s$ represents a driving contribution
to the selected field component, whereas a negative
$g_s$ represents a damping contribution.

In the fluid model, 
linearizing $J_s=q_sn_su_s$ 
gives
the current perturbation
$\delta J_s=q_s(n_s\delta u_s+u_s\delta n_s)$.
For the selected complex frequency $\omega$ at wavenumber $k$,
the linearized fluid equations give the density and velocity
responses per unit electric-field perturbation,
$\delta n_s/\delta E$ and $\delta u_s/\delta E$.
The corresponding current response is
$\delta J_s/\delta E
=q_s(n_s\,\delta u_s/\delta E+u_s\,\delta n_s/\delta E)$.
Applying the same current--field relation as in
\eqrefn{eq:current} gives the theoretical growth-rate
contribution,
$-\operatorname{Re}(\delta J_s/\delta E)/\epsilon_0$.

The field amplitude peaks at 34.15~ns and then decreases
(\figref{fig:decay_detail}(b)). Table~\ref{tab:balance} gives the
contribution of each population to the field-amplitude growth rate
over the 37.64--49.07~ns decay interval. For each population, we
integrate its instantaneous rate $g_s$ from the start of the interval
($t_0=37.64$~ns) to each successive particle-output time $t$.
The integral $\int_{t_0}^{t} g_s(t')\,dt'$ gives that population's
contribution to the change in log amplitude between $t_0$ and $t$.
We fit a straight line to these values as a function of time and
report its slope in the table.

In PIC, background-electron absorption
($-2.331\times10^8~\mathrm{s}^{-1}$) exceeds emitted-electron drive
($+2.006\times10^8~\mathrm{s}^{-1}$) in magnitude. 
The population contributions 
sum to 
$-0.3255\times10^8~\mathrm{s}^{-1}$.
This differs by less than 0.8\% from the rate obtained by fitting
the PIC field amplitude ($-0.3279\times10^8~\mathrm{s}^{-1}$).
M3 also predicts net decay, but both
background-electron absorption ($-0.398\times10^8~\mathrm{s}^{-1}$)
and emitted-electron drive ($+0.047\times10^8~\mathrm{s}^{-1}$)
are much weaker. The two large electron contributions in PIC largely
cancel, which explains how PIC and M3 can give similar net decay
rates despite their different population responses.

To separate the density and velocity variations contributing
to each population current, we define the regional mean density
$\bar n_s$ and density-weighted mean drift velocity $\bar u_s$
over $z_1=0.5$--$z_2=9.5$~cm:

\begin{equation}
\label{eq:regionalmeans}
\begin{aligned}
\bar n_s &=
\frac{1}{\ell}\int_{z_1}^{z_2} n_s(z)\,dz,\\
\bar u_s &=
\frac{\displaystyle\int_{z_1}^{z_2}n_s(z)u_s(z)\,dz}
{\displaystyle\int_{z_1}^{z_2}n_s(z)\,dz},
\qquad \ell=z_2-z_1.
\end{aligned}
\end{equation}

With this definition, $q_s\bar n_s\bar u_s$
is the regional mean current density.

Writing $n_s=\bar n_s+n'_s$ and
$u_s=\bar u_s+u'_s$ in the definition
$J_s=q_sn_su_s$ gives

\begin{equation}
\label{eq:decomposition}
J_s=q_s\bigl(
\bar n_s\bar u_s
+\bar u_sn'_s
+\bar n_su'_s
+n'_su'_s
\bigr).
\end{equation}

The second and third terms describe current variations
associated with density and velocity, respectively,
while the fourth term represents their cross contribution.
We project these three terms separately using the same
mean-subtracted Hann projection and apply \eqrefn{eq:current},
with $J_s$ replaced by the corresponding current term,
to obtain their contributions to the field-amplitude
growth rate. The spatially uniform first term is removed
by the mean subtraction.

Fast emitted-electron density modulation supplies the main drive,
while background-electron velocity modulation supplies the main absorption.

For this current decomposition, fast and slow emitted electrons
are separated using a fixed boundary
$v_{\mathrm{cut}}=2.1966\times10^6\units{m\,s^{-1}}$.
This value is the median regional peak midpoint during
64.531--88.058~ns in the same trajectory.
Keeping the boundary fixed maintains consistent fast--slow
population labels throughout the current analysis.
In Section~\ref{sec:growth}, the local peak midpoint is instead
identified at each axial position to track the evolving
two-peak region.

The pressure response provides a further test of the fluid approximation.
M1 and M3 both require pressure and density perturbations to
be positively proportional, with coincident oscillation peaks and troughs.
Axial PIC pressure is calculated from the central second velocity
moment. Pressure and density perturbations are projected at the
field wavenumber. Over the common decay interval, a linear background
and oscillating components are fitted simultaneously, using the
frequency and growth rate obtained from the field. The phases of the
oscillating components are then compared; zero phase difference
denotes an in-phase response.

The measured pressure--density phase differences are
$84.3^\circ$ for all emitted electrons and $-19.9^\circ$
for background electrons. The large phase shift of the
emitted-electron pressure response shows that it is not
in phase with density. The associated pressure-gradient
force therefore changes the phase of the momentum and
current responses relative to the M3 closure.
This phase difference helps explain the weaker
emitted-electron drive predicted by M3 in
Table~\ref{tab:balance}.

\subsection{Early growth while the two-peak region expands}
\label{sec:growth}

The preceding analysis examined wave decay and population response at intermediate neutral densities. At the lowest density, the fast and slow peaks in the emitted-electron distribution are clearly separated, and the region containing both peaks expands during early growth.

We examine $\NHe=5\times10^{19}\units{m^{-3}}$ over 13.444--16.133 ns, a continuous early-growth interval with an approximately constant field log-amplitude slope and synchronized particle records. The source-side region 0.5--2.0 cm covers the early development of the two peaks. Velocity histograms in 0.1 mm axial bins are smoothed to suppress sampling noise, and the fast and slow peaks and their midpoint $v_{\mathrm{cut}}$ are identified locally. This bin width resolves changes in peak position across the region. The longest continuous two-peak region defines its length $L(t)$; the overlap of these regions across the selected times defines a fixed spatial background for theory.

\begin{figure}[!tb]
\centering
\includegraphics[width=\linewidth]{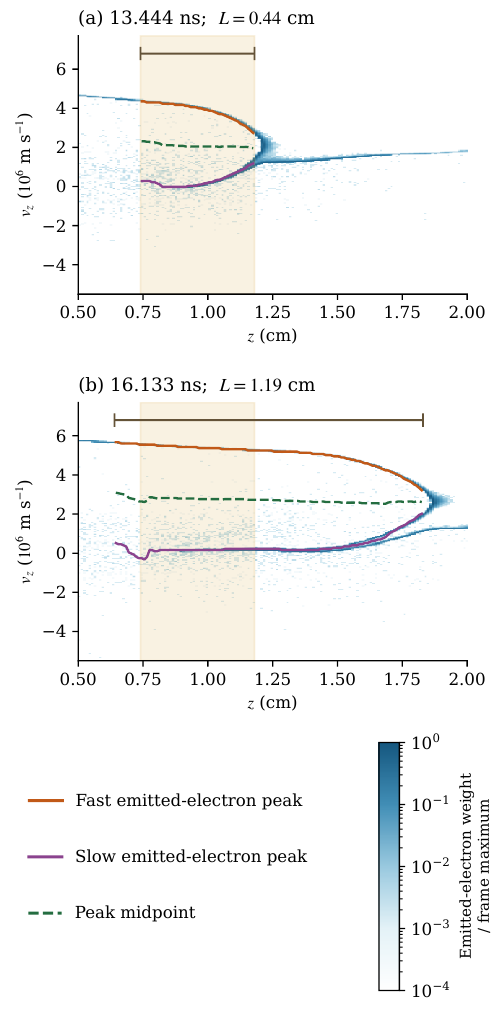}
\caption{Spatial expansion at $\NHe=5\times10^{19}\units{m^{-3}}$. (a,b) Emitted-electron distributions, with darker blue indicating greater particle weight normalized to each frame maximum. Orange, purple and green lines mark the fast peak, slow peak and their midpoint $v_{\mathrm{cut}}$. Brackets mark the two-peak length $L$; shading marks the common theory region.}
\label{fig:expansion}
\end{figure}

\begin{figure}[!tb]
\centering
\includegraphics[width=\linewidth]{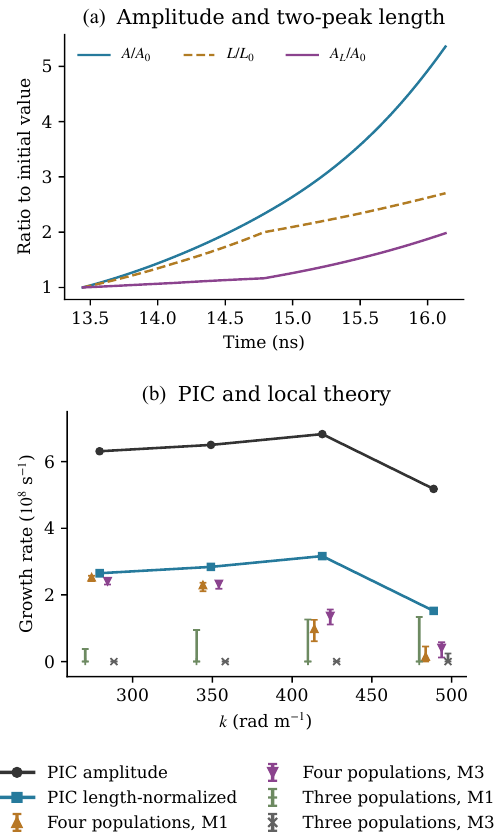}
\caption{Amplitude growth and local theory at $\NHe=5\times10^{19}\units{m^{-3}}$. (a) Amplitude, two-peak length and length-normalized amplitude relative to their values at 13.444 ns. (b) Original and length-normalized PIC growth rates fitted separately for each wavenumber component, compared with three- and four-population theory. Theory markers and bars give spatial medians and full ranges.}
\label{fig:growth_comparison}
\end{figure}

Fast and slow peaks coexist at the same axial positions (\figref{fig:expansion}(a,b)), consistent with \figref{fig:distribution}. The continuous two-peak region grows in length from 0.44 to 1.19 cm. The interval 0.74--1.18 cm is shared by all five particle frames and is used for the local theoretical backgrounds, so the fast and slow populations are sampled at the same positions throughout. The field spectrum still uses the full 0.5--9.5 cm interior region. The local fluid analysis therefore uses the source-side interval where both peaks coexist, whereas the PIC field amplitude is projected over the wider 0.5--9.5 cm interior region.

For each model and pressure closure, the largest growth rate among all dispersion roots is selected in each spatial bin and particle frame. These rates are trapezoidally averaged over the five frames. Figure~\ref{fig:growth_comparison}(b) reports the spatial median and full range over 0.74--1.18 cm. Unlike the decay analysis, this procedure compares the largest local growth available in the model rather than a branch identified by a measured oscillation frequency.

The temporal variance of each field coefficient over 0--134.440 ns defines its fluctuation power. The contiguous wavenumber band with fluctuation power above half its maximum spans $k_4$--$k_7$. We combine the amplitudes in this band as $A(t)=[\sum_{m=4}^{7}A_{k_m}^2(t)]^{1/2}$. Relative to $t_0=13.444\units{ns}$, this amplitude increases about 5.36-fold while $L$ increases about 2.7-fold. To examine the amplitude increase relative to that expansion, define
\begin{equation}
\label{eq:length}
\begin{aligned}
A_L(t)&=A(t)\frac{L(t_0)}{L(t)},\\
\frac{A_L(t)}{A(t_0)}&=\frac{A(t)/A(t_0)}{L(t)/L(t_0)},\\
A_{k_m,L}(t)&=A_{k_m}(t)\frac{L(t_0)}{L(t)},
\qquad m=4,\ldots,7,\\
\gamma_{k_m,L}
&=\frac{d\ln A_{k_m,L}}{dt}
=\gamma_{k_m}-\frac{d\ln L}{dt}.
\end{aligned}
\end{equation}
The normalization takes $A\propto L$ as a reference: if the measured amplitude grew only in proportion to the two-peak length, $A_L$ would remain constant. Instead, $A_L$ increases by about 1.98-fold (\figref{fig:growth_comparison}(a)).

Particle and field records are saved every 100 and 10 steps, respectively, giving five particle times and 41 field times in this interval. To evaluate $A_L$ at the field times, $\ln L$ is linearly interpolated between particle records. The same fit times and weights are used for both amplitudes, so the slope of $\ln A_L$ equals the slope of $\ln A$ minus the slope of $\ln L$. The fitted growth rate decreases from $6.16\times10^8\units{s^{-1}}$ for $A$ to $2.51\times10^8\units{s^{-1}}$ for $A_L$.

For Figure~\ref{fig:growth_comparison}(b), the PIC growth rates are fitted separately for each wavenumber component. The original rate $\gamma_{k_m}$ is obtained from a linear fit to $\ln A_{k_m}(t)$, and the length-normalized rate $\gamma_{k_m,L}$ from a fit to $\ln A_{k_m,L}(t)$, using the same 41 field-output times over 13.444--16.133 ns. The length $L(t)$ is obtained from the particle records, with $\ln L$ linearly interpolated to the field-output times. The values plotted in panel (b) therefore describe the individual $k_m$ components, whereas the 5.36-fold and 1.98-fold amplitude changes in panel (a) refer to the combined $k_4$--$k_7$ band.

To check the particle-output cadence, we also analyse a 41-frame record saved every 10 steps over the same interval. Keeping every tenth frame changes the length-normalized growth rate by 4.3\% relative to using all 41 frames. The 100-step cadence therefore resolves the change in $L(t)$ adequately for this comparison.

The four-population models retain substantial electron-scale growth, whereas the three-population spatial medians are near zero. At $k_4$, the length-normalized PIC rate is $2.65\times10^8\units{s^{-1}}$, compared with four-population medians of $2.53$ and $2.41\times10^8\units{s^{-1}}$ for M1 and M3. This contrast shows that retaining the separate fast and slow emitted-electron drifts is important for capturing the local growth predicted at low density.

Across $k_4$--$k_7$, PIC growth peaks at $k_6$, whereas the theoretical spatial medians decrease with wavenumber; the discrepancy therefore grows toward higher $k$. The four-population model captures the low-wavenumber growth scale, while its predicted wavenumber trend differs from PIC. The early-growth interval spans less than half an oscillation cycle, so we compare amplitude-growth rates rather than track a frequency-matched branch. The next subsection examines how population splitting affects the high-density response and how finite propagation distance limits the amplification of local growth.

\subsection{High-density response and finite propagation distance}
\label{sec:spatial}

The preceding low-density analysis showed that resolving fast and slow emitted electrons improves the local growth-rate comparison. At the high-density end, a split emitted-electron distribution can still affect the local theoretical response even though the large-scale PIC structures are smoother. We use $\NHe=5\times10^{21}\units{m^{-3}}$, a factor of 100 above the low-density case, to compare merged and separate populations under contrasting collisional conditions.

The saved particle records over 0--134.440 ns are searched within 0.5--2.0 cm for continuously resolved fast and slow peaks with common spatial support. The longest such interval is 27.560--34.954 ns. Its 12 particle times share the region 0.56--0.64 cm, which is held fixed for both models. This construction compares the effect of population representation using identical positions and particle backgrounds. The first discrete spectral component on the 0.5--9.5 cm interior grid gives $k_1=69.821\units{rad\,m^{-1}}$. The 0.56--0.64 cm interval is used to obtain the local particle moments for the dispersion analysis.

\begin{figure*}[!tb]
\centering
\includegraphics[width=\textwidth]{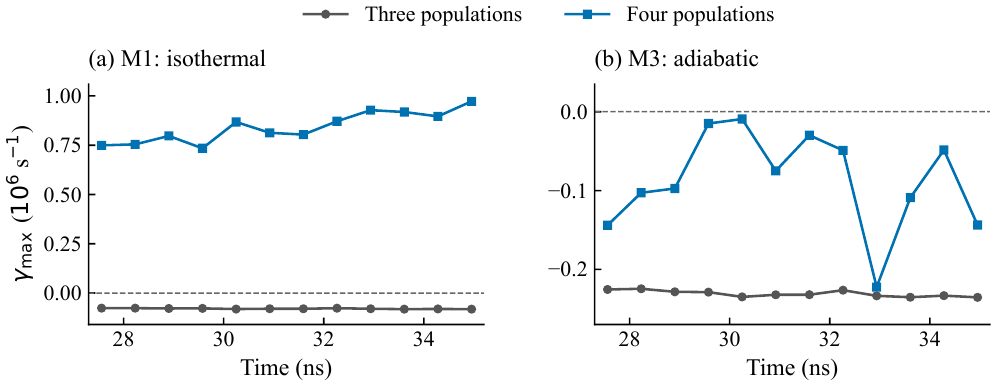}
\caption{Maximum theoretical rates for the high-density three- and four-population models, using the backgrounds and wavenumber in \tabref{tab:high}. (a) Isothermal M1; (b) adiabatic M3. Grey circles denote the merged model and blue squares the full four-population model. Markers identify particle-output times, connecting lines guide the eye and dashed horizontal lines mark zero growth. The panels use different vertical-axis ranges.}
\label{fig:high}
\end{figure*}

At each particle time, all six three-population roots and all eight four-population roots are evaluated for each closure. The maximum rate is $\gamma_{\max}=\max_j\operatorname{Im}\omega_j$. A positive value identifies a growing branch at the selected $k$ in that background; a negative value means all roots at that $k$ decay. Table~\ref{tab:high} gives the range over the twelve times, and \figref{fig:high} shows their variation.

\begin{table*}[!tb]
\caption{Maximum theoretical rate $\gamma_{\max}$ for the same high-density split-electron background, in $10^6\units{s^{-1}}$. Ranges cover twelve times over 27.560--34.954 ns, the region 0.56--0.64 cm and $k_1=69.821\units{rad\,m^{-1}}$.}
\label{tab:high}
\centering
\small
\begin{tabular}{lcc}
\toprule
Model & M1: isothermal & M3: adiabatic\\
\midrule
Merged three populations & $[-0.082,-0.077]$ & $[-0.236,-0.225]$\\
Full four populations & $[+0.734,+0.973]$ & $[-0.223,-0.009]$\\
\bottomrule
\end{tabular}
\end{table*}

With M1 (\figref{fig:high}(a)), the merged model predicts decay, while the four-population model gives weak growth, $\gamma_{\max}=0.734$--$0.973\times10^6\units{s^{-1}}$. With M3 (\figref{fig:high}(b)), all roots of both models decay at this wavenumber. Population splitting and pressure closure therefore jointly affect the predicted local stability.

The next question is whether a growing local response has enough distance to amplify appreciably inside the gap. For this calculation, the same high-density split-population backgrounds and wavenumber are used to follow one electron branch of the four-population model continuously. Its phase velocity lies between the fast- and background-electron drifts. Its current response shows that fast emitted electrons supply energy to the wave, whereas slow emitted and background electrons absorb it. Slow emitted electrons account for 83.1--87.9\% of the emitted-electron density in this region, measured as $n_{\mathrm{slow}}/(n_{\mathrm{fast}}+n_{\mathrm{slow}})$.

To obtain a characteristic spatial scale, each source-side background is held fixed and extended uniformly along a 10 cm path. For weak temporal growth or damping, the group velocity $v_g$ relates temporal and spatial rates~\cite{r34}:
\begin{equation}
\label{eq:spatial}
\begin{aligned}
v_g&=\frac{\partial\omega_r}{\partial k},\qquad
\alpha\simeq\frac{\gamma}{|v_g|},\\
L_g&=\frac{1}{\alpha}\quad(\alpha>0),\qquad
\frac{A(\ell)}{A(0)}\simeq e^{\alpha\ell}.
\end{aligned}
\end{equation}
The derivative follows the same branch at fixed background. Here $\ell$ is propagation distance and $L_g$ is the distance for an $e$-fold amplitude increase.

For the high-density branch, the group speed is $1.62$--$1.71\times10^6\units{m\,s^{-1}}$ and $|\gamma/\omega_r|<0.0084$, supporting the weak-growth conversion. M1 gives only 4.4--6.0\% amplitude growth over 10 cm. M3 gives slight decay, with amplitude factors of 0.9775--0.9995.

For $\NHe=5\times10^{19}\units{m^{-3}}$, the four-population calculation uses the same five particle states and 0.74--1.18 cm region as the expansion analysis, with temporal reference $k_4=279.283\units{rad\,m^{-1}}$. The fastest-growing temporal root is selected at each time for each closure. Since $|\gamma/\omega_r|=0.19$--0.29, the weak-growth conversion is not used. Instead, a spatial root is obtained directly from $D(\omega_r,k)=0$ at real frequency, by continuation from the selected temporal root. For forward propagation, the spatial growth rate is $\alpha=-\operatorname{Im}k$.

\begin{table*}[!tb]
\caption{Spatial response of the four-population model in fixed, uniform backgrounds. The growth length $L_g$ gives an $e$-fold amplitude increase; a dash denotes a decaying branch.}
\label{tab:lengths}
\centering
\small
\begin{tabular}{cllc}
\toprule
$\NHe\ (10^{21}\units{m^{-3}})$ &
Pressure closure &
Response &
$L_g$ (cm)\\
\midrule
0.05 & M1: isothermal & Growth & 0.894--1.141\\
0.05 & M3: adiabatic & Growth & 1.083--1.245\\
5 & M1: isothermal & Weak growth & 172.3--233.4\\
5 & M3: adiabatic & Decay & ---\\
\bottomrule
\end{tabular}
\end{table*}

Table~\ref{tab:lengths} shows a clear contrast between the two density limits. At low density, the growth length of 0.894--1.245 cm is comparable to the 0.44--1.19 cm extent of the two-peak region. At high density, the M1 growth length of 172.3--233.4 cm (1.72--2.33 m) is much greater than the 10 cm gap, consistent with the small 4.4--6.0\% amplitude increase; M3 predicts decay. The local response must therefore be interpreted relative to the distance available for amplification: the low-density growth length is on the scale of the two-peak region, whereas the high-density growth length greatly exceeds the electrode spacing.

These estimates place the local dispersion results in the context of the finite ETC gap. Comparing the growth lengths with the two-peak region and electrode spacing identifies the characteristic distance over which the modeled disturbances can amplify under the sampled background conditions. Population separation and pressure closure determine the local wave response, while the available propagation distance sets the scale of its cumulative amplification. Section 5 then examines late-time anode collection and current variability under the same prescribed conditions.

\section{Late-time electron transport and implications for cooling}
\label{sec:transport}

The wave analysis returns us to the engineering question:
how efficiently are emitted electrons transported to the anode
in the fluctuating states?
The late-time phase-space patterns in \figref{fig:late}
show density-dependent differences in electron dynamics.
We therefore compare mean collection and current fluctuations
across the density scan.

All late-time statistics use 6.8--11.2 $\mu$s.
The mean anode current varies modestly within this interval,
and checks over neighbouring windows preserve the main
transport differences.
The common interval therefore supports comparison of the
transport level and its temporal variation.

The collected current follows the emitted-electron flux
definition of the full-gap transport study~\cite{r14}.
Particle positions and velocities predict the number $N_a$
of emitted macroparticles crossing the anode in the next step,
giving
$J_a=ew_pN_a/\Delta t$,
where $w_p$ is the macroparticle weight per unit transverse area.
The mean anode collection efficiency is
$\eta_a=100\%\times\overline{J_a}/J_{\mathrm{emit}}$,
with $J_{\mathrm{emit}}=e\Gamma_{\mathrm{emit}}$
and the average taken over all time-step samples in the interval.

For relative fluctuations, the current is first averaged over
consecutive, nonoverlapping groups of 100 steps
(0.6722 ns).
The resulting block means are denoted $J_{a,100}$.
Their coefficient of variation is
\begin{equation}
\label{eq:cv}
C_V=100\%\times
\frac{\sigma(J_{a,100})}
{\langle J_{a,100}\rangle}.
\end{equation}

Here, $\sigma$ and $\langle\cdot\rangle$ denote the
standard deviation and mean of the block means, respectively.

Only complete blocks are used, with the same block length
for all densities.
At $\NHe=5\times10^{19}\units{m^{-3}}$, for example,
6545 complete block means have a mean of
$12.82\units{A\,m^{-2}}$ and a standard deviation of
$6.97\units{A\,m^{-2}}$,
giving $C_V\simeq(6.97/12.82)\times100\%=54.4\%$.

\begin{table}[!tb]
\caption{Emitted-electron anode collection and relative current fluctuations over 6.8--11.2 $\mu$s. The prescribed emission current density is $e\Gamma_{\mathrm{emit}}\simeq12.82\units{A\,m^{-2}}$. Efficiencies use unrounded values; fluctuations use 0.6722 ns block averages.}
\label{tab:transport}
\centering
\small
\setlength{\tabcolsep}{4pt}
\begin{tabular}{@{}crrr@{}}
\toprule
\shortstack{$\NHe$\\$(10^{21}\units{m^{-3}})$} &
\shortstack{$\overline{J_a}$\\$(\mathrm{A\,m^{-2}})$} &
\shortstack{$\eta_a$\\(\%)} &
\shortstack{$C_V$\\(\%)}\\
\midrule
0.05 & 12.82 & 100.0 & 54.4\\
0.1  & 12.81 & 99.9 & 52.5\\
0.5  & 12.77 & 99.6 & 48.5\\
1    & 12.66 & 98.7 & 36.6\\
1.5  & 12.58 & 98.1 & 1.9\\
2    & 12.46 & 97.2 & 1.5\\
2.9  & 12.23 & 95.4 & 1.5\\
5    & 11.50 & 89.7 & 2.0\\
\bottomrule
\end{tabular}
\end{table}

The three lowest densities retain 99.6--100.0\% mean
collection despite 48.5--54.4\% relative fluctuations
(\tabref{tab:transport}).
Strong time variation therefore coexists with efficient
average transport of the emitted electrons.

The sharpest change occurs between
$10^{21}$ and $1.5\times10^{21}\units{m^{-3}}$:
$C_V$ decreases from 36.6\% to 1.9\%,
while the collection efficiency changes only
from 98.7\% to 98.1\%.
At higher densities, $C_V$ remains between
1.5\% and 2.0\%, whereas the collection efficiency
decreases from 98.1\% to 89.7\%.
The two measures therefore respond to neutral density
at different rates.

Section 4 resolves the population contributions to
electrostatic-wave growth and decay over nanosecond
time intervals.
The present statistics quantify late-time anode collection
over 6.8--11.2 $\mu$s, with current variability calculated
from 0.6722 ns block means.
Together, these analyses describe complementary aspects
of the same density-dependent plasma system:
wave dynamics and late-time electron transport.

For ETC, the practical implication is that fluctuation
strength and mean electron collection should both be
examined~\cite{r5,r14}.
In this model, a strongly fluctuating state can still collect
almost all emitted electrons, while a smoother state can
collect a smaller fraction.

Mean collection efficiency and $C_V$ therefore provide
complementary information for ETC operation:
the former quantifies the fraction of emitted current
collected at the anode, while the latter quantifies
the temporal variability of that current at the selected
averaging scale.
In the present helium gap, these trends are obtained
with prescribed emission, pair replenishment, a fixed
40 V bias, and the specified elastic and charge-exchange
collision channels.
Combining the transport statistics with the energy balances
of the emitting surface and circuit provides a system-level
basis for relating plasma transport to heat removal
and electrical energy transfer.

\section{Conclusions}

Electron transpiration cooling depends on the transport of thermionically emitted electrons through the surrounding plasma. This study used PIC--MCC simulations of a finite helium gap to examine the collective dynamics accompanying that transport. Varying the neutral density reveals changes in electron velocity structures and electrostatic activity, together with distinct changes in mean collection and current variability.

In the analysed decay stage, fast emitted-electron density modulation supplies the main drive, while background-electron velocity modulation supplies the main absorption. Overall, absorption exceeds drive and the wave decays. The adiabatic three-population model captures the frequency scale and decay direction across four intermediate-density cases, but greatly underestimates the separate driving and absorbing responses. A close net rate can therefore conceal errors in the underlying population response. The measured pressure--density phases identify one limitation of the in-phase fluid closure.

At low density, retaining fast and slow emitted electrons yields substantial local fluid growth that is largely lost when they are merged. Normalizing the PIC amplitude by the expanding two-peak length improves the low-wavenumber comparison, while higher-wavenumber discrepancies remain. At high density, population splitting and pressure closure affect the predicted local stability. Fixed, uniform backgrounds give centimetre-scale growth lengths at low density and a metre-scale isothermal growth length at high density; the high-density adiabatic branch decays. Interpreting a local growth rate therefore requires considering how the background evolves and how far the disturbance can propagate.

For ETC, the late-time results show that strong fluctuations can coexist with efficient mean transport: the three lowest densities retain 99.6--100.0\% mean anode collection, whereas the highest density has both weaker fluctuations and lower collection. Within this model, a quieter current does not necessarily indicate better average transport. The early wave analysis and late current statistics characterize different stages, without establishing a direct link between individual modes and later fluctuations. Engineering assessment should therefore consider mean electron collection and current variability together, and relate them to the surface and circuit energy balance before drawing conclusions about cooling performance.

\section*{Acknowledgment}

This work was supported by the National Natural Science Foundation of China (Grant Nos.~U23B2033 and 52472403).

\section*{Data Availability}

The data that support the findings of this study are available
from the corresponding authors upon reasonable request.

\end{document}